\documentclass{article}

\usepackage{geometry}
\usepackage{mathpazo}
\usepackage{amssymb}
\usepackage[fleqn]{amsmath}
\usepackage{graphicx}
\usepackage{epstopdf}
\usepackage[titletoc,title]{appendix}
\usepackage{gensymb}
\usepackage{bm}
\usepackage{authblk}
\usepackage{xcolor}

\author[1]{J\'er\'emy R. Rouxel}
\author[2]{Riccardo Mincigrucci}
\author[2]{Danny Fainozzi}
\author[2]{Claudio Masciovecchio}

\affil[1]{Chemical Sciences and Engineering Division, Argonne National Laboratory, Lemont, Illinois 60439, United States}
\affil[2]{Elettra-Sincrotrone Trieste, SS 14 - km 163.5, 34149 Basovizza, Trieste, Italy}

\title{X-ray Circular Dichroism measured by cross-polarization X-ray Transient Grating}

\date{\today}

\begin{document}

\maketitle

\begin{abstract}
Measuring natural circular dichroism in the X-ray regime to extract stereochemical information from chiral molecules in solution remains a challenge.
This is primarily due to technical limitations of the existing synchrotron sources. 
It hinders access to measurements of local chirality by exploiting core hole electronic transitions. 
In response to this challenge, we propose an alternative approach: utilizing XFEL-based cross-polarization X-ray Transient Grating (XTG). 
This method provides an indirect means to measure X-ray Circular Dichroism (XCD). 
Notably, our findings reveal that the signal only emerges once the excited cores have undergone dephasing through relaxation.
XTG is now routinely measured in the XUV regime, and has recently been made available for hard X-rays.
Free electron lasers now offer polarization controls and XTG can be extended to various polarization states for the two pump beams, making XCD measured by XTG feasible at the current state of the art.
\end{abstract}

\section{Introduction}

The exploration of chiral molecules has been captivating since the pioneering studies conducted by Pasteur in 1848\cite{pasteur1848relations}. 
Since these initial investigations, chirality has held a central role in the field of material sciences and chemistry.
Optical Circular Dichroism (CD) spectroscopy stands out as the predominant method for the detection and study of enantiomers\cite{berova2000circular}, i.e. molecules with opposite chirality. This technique relies on the measurement of the difference in absorption exhibited by left and right circularly polarized light. Enantiomers manifest distinct responses to these two forms of light, absorbing them differently.
While optical CD is a powerful tool for studying the global chirality of molecules, it encounters limitations in accessing local chirality\cite{rouxel2022molecular}. Indeed, optical CD primarily relies on transitions involving outer-shell electrons, providing information about the overall molecular structure and chiral properties. 

In contrast, X-ray CD, exploiting core hole electronic transitions, offer a more in-depth understanding of the electronic and structural aspects of chiral molecules\cite{rouxel2016non}. This capability enables the examination of local chirality, providing insights into the specific arrangement of the electron density around the chiral centers.
Mirror symmetry breaking can occur at specific sites within molecules, at stereogenic centers, or be delocalized everywhere like in helical compounds\cite{freixas2023x}. 
Optical CD cannot discriminate these types of chiralities and their localization within molecules.
Alternatively, a chiral molecule can provide a vanishing X-ray CD if the probed atom is located far enough from chiral features and is insensitive to them.
Despite the significance of natural X-ray Circular Dichroism (XCD), successful experiments have been relatively few, typically limited to crystals or the gas phase\cite{alagna1998x,goulon1998x,platunov2021x}. 
The primary challenge lies in the subtle difference in signal arising from the differential absorption of left- and right-polarized photons.
As a consequence, the experimental detection of natural XCD, especially in more complex environments like solutions, has proven to be a formidable task.

Addressing this challenge requires innovative approaches and advancements in experimental techniques to enhance the signal-to-noise ratio and improve the sensitivity of XCD measurements. Overcoming these limitations is crucial for unlocking the full potential of XCD in providing detailed insights into the structural and electronic features of chiral molecules, which is essential for various scientific disciplines, including chemistry, biochemistry, and materials science\cite{kasprzyk2010pharmacologically,smith2009chiral,wang2016optical,cline2005physical}.
In the present work we theoretically demonstrate that cross-polarization X-ray Transient Grating (XTG) provides an indirect method for measuring X-ray CD.

XTG is a four-wave-mixing(4WM) technique in which two coincidence pump beams with the same spectral profile are crossed with an angle $2\theta$ on a sample and create a transient interference pattern of excitation\cite{brown1999femtosecond}. 
The diffraction of a delayed probe emitted in a background free direction\cite{fourkas1992transient} is then detected.
Similar to standard pump-probe experiments, the diffracted intensity contains information about ultrafast relaxation pathways.
Additionally, the transient grating signal can also decay by transport of the excited carriers, which is then monitored by the XTG techniques\cite{johnson2013direct}. 
TG has been commonly employed since the 80s in optical laboratories, and the work of Bencivenga et al.\cite{bencivenga2015four, bencivenga2019nanoscale, maznev2018generation} at the FEL FERMI has made TG experiments possible in the XUV using the beam crossing geometry in the last decade.
Recently, the use of diffractive optics has opened it in the hard X-ray regime\cite{rouxel2021hard}.

As a 4WM technique, XTG can be described by third-order nonlinear polarizability $P^{(3)}_\text{XTG}$, itself containing a convolution between the four-point matter correlation functions and the incoming field time envelopes. 
However, if the delay $\tau$ between the two pumps and the probe is longer than the matter electronic decoherence time, the matter correlation function can be factorized into two two-point correlation functions.
This is typically the case when thermal transport is investigated and the transport during the delay $\tau$ is essentially classical.
In this case, the transient excitation grating is used to deposit energy, in the form of heat after relaxation, and quantum features are lost.
While it may be detrimental to measure, e.g. core-hole relaxation using XTG, this can be used as an indirect measure of linear responses, including the chiral linear response.

In the optical regime, the idea originates from foundational works of Fayer\cite{fourkas1992gratingI,fourkas1992gratingII} and Terazima\cite{terazima1995new,terazima1996transient}.
Using crossed polarization states, they showed that the polarization grating can be converted into an intensity grating if the excited matter possesses chirality, i.e. a differential absorption of left and right circular polarization of light (CPL).
In the following, we show how this allows the measure of X-ray Circular Dichroism (XCD) indirectly.
In the optical regime, the widespread availability of commercially available spectrometer for CD has hindered the development of this more complicated, indirect measurement.
Additionally, homodyne-detected signals are not sensitive to sign changes in the signal amplitudes.
As a consequence, XTG-detected XCD cannot distinguish opposite enantiomers.
It is still sensitive to chirality in the sense that the signal vanishes in achiral systems, but is not enantioselective. 
However, in the X-rays or EUV, XCD is more challenging due to the difficulty to obtain left and right CPL with high stability and polarization purity.
For example, FERMI already offers\cite{roussel2017polarization, allaria2014control} the required cross-polarization schemes, making XTG-measured XCD readily available.

In this work, we first review general signal expressions of the XTG signal. Second, we discuss the geometry of the cross-polarization XTG. Third, we show how the rotatory strength emerges from dephased XTG signal. Finally, a simulation of XTG-measured XCD from L-tryptophan is made as an illustration.

\section{Signal definition}

\begin{figure}[h]
    \centering
    \includegraphics[width=0.4\textwidth]{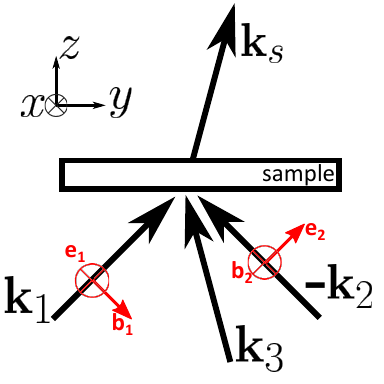}
    \caption{Geometry of cross-polarization XTG setup. The polarizations $\bm e_1$ and $\bm e_2$ of pulses $\bm E_1$ and $\bm E_2$ are orthogonal to each other. The corresponding polarizations $\bm b_1$ and $\bm b_2$ of the magnetic fields $\bm B_1$ and $\bm B_2$ are also shown. The TG signal is emitted in the phase-matching direction $\bm k_s = \bm k_1 -\bm k_2 + \bm k_3$.}
    \label{fig:setup}
\end{figure}

The XTG signal is a spontaneous, coherent signal and can thus be expressed as the amplitude squared of the nonlinear matter polarization\cite{mukamel1995principles}:
\begin{equation}
S(\tau) = \int dt |P_\text{XTG}^{(3)}(t,\tau)|^2
\end{equation}
\noindent where $P_\text{XTG}^{(3)}$ is the third-order matter polarization emitting a spontaneous field in the $\bm k_s = \bm k_1 - \bm k_2 + \bm k_3$ phase-matching direction, see Fig.\ref{fig:setup}. 
$\tau$ is the delay between the EUV pumps and the probe pulse and $t$ is the integration time of the detector.
The nonlinear polarization $P_\text{XTG}^{(3)}$ is generated by perturbative interactions with two coincident, non-collinear pump pulses $\bm E_1$ and $\bm E_2$ and a delayed diffracted probe $\bm E_3$. 
As shown in Fig. \ref{fig:setup}, pulses $\bm E_1$ and $\bm E_2$ are cross-polarized (red marks in the scheme).
The probe polarization is unimportant for the proposed homodyne-detected XTG and is thus left unspecified. 
At FELs, $\bm E_1$ and $\bm E_2$ are high energy pulses (EUV or X-ray), while $\bm E_3$ is either optical or high frequency depending on the periodicity of the generated transient grating.
Full X-ray (pumps and probe) XTG is still a challenging technique to implement\cite{rouxel2021hard}.

Our derivation for XTG-detected XCD signal is based on the multipolar field-matter interaction Hamiltonian truncated at the magnetic dipole order\cite{craig1998molecular}:
\begin{equation}
H_\text{int}(t) = -\bm\mu\cdot \bm E(t) -\bm m\cdot \bm B(t)
\end{equation}
\noindent where $\bm \mu$ and $\bm m$ are the electric and magnetic transition dipoles respectively and $\bm E$ and $\bm B$ are the incoming electric and magnetic field amplitudes.
We do not consider electric quadrupole coupling in the following since it vanishes upon rotational averaging over random sample orientation\cite{craig1998molecular}.

\begin{figure}[h]
    \centering
    \includegraphics[width=0.9\textwidth]{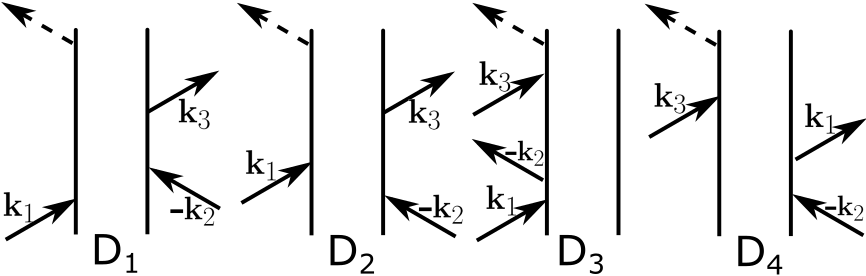}
    \caption{Ladder diagrams contributing to the XTG signal. Interactions with $\bm k_1$ and $\bm k_2$ can be either with electric dipole or magnetic dipole coupling.}
    \label{fig:diag1}
\end{figure}

The polarization $P_\text{XTG}^{(3)}$ is given by a summation over the ladder diagrams relevant for the XTG technique\cite{mukamel1995principles} and shown in Fig. \ref{fig:diag1}:
\begin{equation}
P_\text{XTG}^{(3)}(t, \tau) = \sum_i P_{D_i}(t, \tau)
\label{eq:sumPdi}
\end{equation}
with
\begin{multline}
P_{D_1}(t, \tau) = \Big(-\frac{i}{\hbar}\Big)^3\int dt_3 dt_2dt_1
\langle \bm \mu_L \mathcal G(t_3)\bm \mu_R \mathcal G(t_2)\bm \mu_R \mathcal G(t_1)\bm \mu_L\rangle 
\bm E_3(t-t_3)\bm E_2^*(t-t_3-t_2+\tau)\bm E_1(t-t_3-t_2-t_1+\tau)
\label{eq:d1}
\end{multline}
\begin{multline}
P_{D_2}(t, \tau) = \Big(-\frac{i}{\hbar}\Big)^3\int dt_3 dt_2dt_1
\langle \bm \mu_L \mathcal G(t_3)\bm \mu_R \mathcal G(t_2)\bm \mu_L \mathcal G(t_1)\bm \mu_R\rangle 
\bm E_3(t-t_3)\bm E_1(t-t_3-t_2+\tau)\bm E_2^*(t-t_3-t_2-t_1+\tau)
\end{multline}
\begin{multline}
P_{D_3}(t, \tau) = \Big(-\frac{i}{\hbar}\Big)^3\int dt_3 dt_2dt_1
\langle \bm \mu_L \mathcal G(t_3)\bm \mu_L \mathcal G(t_2)\bm \mu_L \mathcal G(t_1)\bm \mu_L\rangle 
\bm E_3(t-t_3)\bm E_2^*(t-t_3-t_2+\tau)\bm E_1(t-t_3-t_2-t_1+\tau)
\end{multline}
\begin{multline}
P_{D_4}(t, \tau) = \Big(-\frac{i}{\hbar}\Big)^3\int dt_3 dt_2dt_1
\langle \bm \mu_L \mathcal G(t_3)\bm \mu_L \mathcal G(t_2)\bm \mu_R \mathcal G(t_1)\bm \mu_R\rangle 
\bm E_3(t-t_3)\bm E_1(t-t_3-t_2+\tau)\bm E_2^*(t-t_3-t_2-t_1+\tau)
\label{eq:d4}
\end{multline}
\noindent where $\mu_L$ and $\mu_R$ correspond to interaction on the left or the right of the matter density matrix, leading to different phases for different interaction pathways. Eqs.\ref{eq:d1} to \ref{eq:d4} can be read out directly from the diagrams in a systematic manner\cite{mukamel1995principles}.
As a general 4WM technique, the nonlinear polarization is thus given by the convolution of a four-point matter correlation function with the temporal envelopes of the incoming pulses.

In the cross-polarization XTG setup, magnetic dipole interactions with the EUV beams are considered. Each of the 4 diagrams in Fig. \ref{fig:diag1} can be separated into 3 contributions corresponding to different multipolar interaction pathways.
For example, the contributions from diagram $D_1$ in Fig. \ref{fig:diag1} to the signal is now:
\begin{multline}
P_{D_1}^{\mu\mu}(t, \tau) = \Big(-\frac{i}{\hbar}\Big)^3\int dt_3 dt_2dt_1
\langle \bm \mu_L \mathcal G(t_3)\bm \mu_R \mathcal G(t_2)\bm \mu_R \mathcal G(t_1)\bm \mu_L\rangle 
\bm E_3(t-t_3)\bm E_2^*(t-t_3-t_2+\tau)\bm E_1(t-t_3-t_2-t_1+\tau)
\end{multline}
\begin{multline}
P_{D_1}^{\mu m}(t, \tau) = (-\frac{i}{\hbar})^3\int dt_3 dt_2dt_1
\langle \bm \mu_L \mathcal G(t_3)\bm \mu_R \mathcal G(t_2)\bm \mu_R \mathcal G(t_1)\bm m_L\rangle 
\bm E_3(t-t_3)\bm E_2^*(t-t_3-t_2+\tau)\bm B_1(t-t_3-t_2-t_1+\tau)
\label{eq:pd1mum}
\end{multline}
\begin{multline}
P_{D_1}^{m\mu}(t, \tau) = (-\frac{i}{\hbar})^3\int dt_3 dt_2dt_1
\langle \bm \mu_L \mathcal G(t_3)\bm \mu_R \mathcal G(t_2)\bm m_R \mathcal G(t_1)\bm \mu_L\rangle 
\bm E_3(t-t_3)\bm B_2^*(t-t_3-t_2+\tau)\bm E_1(t-t_3-t_2-t_1+\tau)
\end{multline}
\noindent where $P_{D_1}^{\mu\mu}(t, \tau)$ is an electric-dipole only contribution, not sensitive to chirality, and that vanishes when using a cross polarization scheme.
$P_{D_1}^{\mu m}(t, \tau)$ and $P_{D_1}^{m\mu}(t, \tau)$ have one magnetic dipole interaction with either the pulse $\bm E_1$ or $\bm E_2$ respectively.

\begin{figure}[h]
    \centering
    \includegraphics[width=0.5\textwidth]{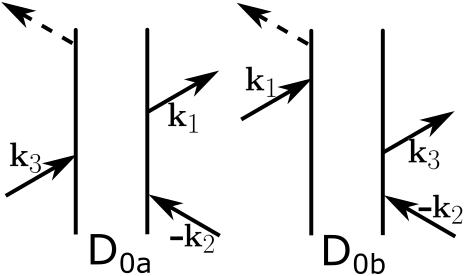}
    \caption{Additional ladder diagrams contributing to the XTG signal when the delay between the EUV pumps and the probe is small.}
    \label{fig:diag2}
\end{figure}

Diagrams in Fig. \ref{fig:diag1} assumes that the pumps and the probe are well separated temporarily, so partial time-ordering of the interactions can be ensured.
Near time zero, when the pumps and the probe temporally overlap, additional diagrams must be taken into account, shown in Fig. \ref{fig:diag2}.
For example, self-diffraction is a degenerate XTG signal with no delayed probe and in which the "probe" interaction is achieved with one of the pump pulses.
This experimentally simpler technique (no delay $\tau$ to implement) would require the addition of these two contributions in the signal description.

\section{Geometry for cross-polarized XTG}

Some geometric identities are needed to describe the signal. We use the frame of coordinates shown in Fig. \ref{fig:setup}. The pulse $\bm E_1$ and $\bm E_2$ are linearly polarized along the $x$ and $y$ axes respectively, and have an angle $\pm\theta$ with respect to the $z$ axis.
Using simple algebra, we can show that the polarizations and wavevectors are:
\begin{eqnarray}
\bm e_1 &=& \bm e_x \nonumber\\
\bm e_2 &=& \cos \theta \bm e_y + \sin\theta \bm e_z \nonumber \\
\bm k_1\cdot \bm r &=&  k_X (y\sin \theta + z\cos \theta) \nonumber\\
\bm k_2\cdot \bm r &=&  k_X (-y\sin \theta + z\cos \theta)
\end{eqnarray}

It will be convenient in the following to use the circular polarization basis defined as:
\begin{equation}
\bm e_L = \frac{1}{\sqrt 2}\begin{pmatrix}1 \\ i \\ 0\end{pmatrix} 
\ \ \ \ \ \
\bm e_R = \frac{1}{\sqrt 2}\begin{pmatrix}1 \\ -i \\ 0\end{pmatrix} 
\end{equation}

We also define a polarization basis for the $\bm B$ field as:
\begin{eqnarray}
\bm b_L = \bm e_z \times \bm e_L = -i \bm e_L\\
\bm b_R = \bm e_z \times \bm e_R = i \bm e_R
\end{eqnarray}

Using the spherical basis, the fields can be written as:
\begin{eqnarray}
\bm E_1 &=& E_x(t)\frac{1}{\sqrt{2}}(\bm e_L + \bm e_R) e^{i k_X (y\sin \theta + z\cos \theta)}\\
\bm E_2 &=& E_x(t)\Big(-i\frac{\cos \theta}{\sqrt{2}}(\bm e_L - \bm e_R) +\bm e_z\sin\theta \Big) e^{i k_X (-y\sin \theta + z\cos \theta)}\\
\bm B_1 &=& \frac{E_x(t)}{c}\Big(\frac{\cos \theta}{\sqrt{2}}(\bm b_L + \bm b_R)
-\sin \theta \bm e_z
\Big) e^{i k_X (y\sin \theta + z\cos \theta)}\\
\bm B_2 &=& \frac{E_x(t)}{c} \frac{1}{\sqrt{2}}(-i)(\bm b_L - \bm b_R) e^{i k_X (-y\sin \theta + z\cos \theta)}
\end{eqnarray}

Finally, we provide some identities involving scalar products used in the following:
\begin{eqnarray}
\bm e_L\cdot \bm e_L^* &=& \bm e_R\cdot \bm e_R^* = 1\nonumber\\
\bm e_L\cdot \bm e_R^* &=& \bm e_R\cdot \bm e_L^* = 0\nonumber\\
\bm e_L\cdot \bm b_L^* &=& \bm e_R^*\cdot \bm b_R = i\nonumber\\
\bm e_R\cdot \bm b_R^* &=& \bm e_L^*\cdot \bm b_L = -i
\end{eqnarray}

\section{XCD measured by dephased cross-polarization XTG}

We now show how the chiral-sensitivity appears in cross-polarization XTG.
To simplify the discussion, we consider the impulsive limit for the 3 incoming pulses: $\bm E_i(t) = \bm E_i \delta(t)$.
The impulsive limit cannot include a complete description of resonant spectral features, which are discussed in Appendix \ref{app:resonance}, but is convenient to show how the polarization algebra eliminates the achiral contribution to the XTG signal.
We also restrict the discussion to diagram $D_1$, since the other three give similar contributions.
In the impulsive limit, $P_{D_1}(t, \tau)$ becomes:
\begin{equation}
P^{\mu\mu}_{D_1}(t, \tau) = (-\frac{i}{\hbar})^3 \langle \bm \mu_L \mathcal G(t)\bm \mu_R \mathcal G(\tau) \bm \mu_R \mathcal G(0)\bm \mu_L\rangle \bm E_3 \bm E_2^* \bm E_1 
\end{equation}

Rigorously, the XTG response involves a four-point correlation function.
However, the core excitation triggered by the EUV pulses are usually short lived (tens of fs, or less) and the signal dephases completely during the pump-probe delay $\tau$.
This implies that the correlation function can be factorized into two two-point correlation functions\cite{rappen1994polarization} and that $\mathcal G(\tau)$ is replaced by a classical propagator $U(\tau)$.
If the excited carriers can move during that time, $U(\tau)$ describes the classical transport that is measured in long delay TG experiments.

The contribution $P_{D_1}(t, \tau)$ becomes:
\begin{equation}
P^{\mu\mu}_{D_1}(t, \tau) = (-\frac{i}{\hbar})^3 
\langle \bm \mu_L \mathcal G(t)\bm \mu_R\rangle
U(\tau)
\langle\bm \mu_R \bm \mu_L\rangle \bm E_3 \bm E_2^* \bm E_1 
\end{equation}

If the sample is constituted of randomly oriented components (for example molecules in liquid phase or isotropic powder form), the correlation functions must be averaged over orientations. This leads to:
\begin{equation}
P^{\mu\mu}_{D_1}(t, \tau) = (-\frac{i}{\hbar})^3 \frac{1}{9}
\langle \bm \mu_L\cdot\bm \mu_R\rangle
U(\tau)
\langle\bm \mu_R\cdot\bm \mu_L\rangle \bm E_3 \bm E_2^*\cdot\bm E_1
\end{equation}

Similarly, for the magnetic components, we obtain:
\begin{eqnarray}
P^{\mu m}_{D_1}(t, \tau) &=& (-\frac{i}{\hbar})^3 \frac{1}{9}
\langle \bm \mu_L\cdot\bm \mu_R\rangle
U(\tau)
\langle\bm \mu_R\cdot\bm m_L\rangle \bm E_3 \bm E_2^*\cdot\bm B_1 \\
P^{m\mu}_{D_1}(t, \tau) &=& (-\frac{i}{\hbar})^3 \frac{1}{9}
\langle \bm \mu_L\cdot\bm \mu_R\rangle
U(\tau)
\langle\bm m_R\cdot\bm \mu_L\rangle \bm E_3 \bm B_2^*\cdot\bm E_1 
\end{eqnarray}

Using the identities of the previous section, we can finally calculate the various dot products as follow:
\begin{eqnarray}
\bm E_1\cdot \bm E_2^* &=& -i \frac{E_X^2}{2}\cos \theta e^{2i k_X y \sin \theta} (\bm e_L + \bm e_R)\cdot (\bm e_L - \bm e_R)= 0\\
\bm E_1\cdot \bm B_2^* &=& \frac{E_X^2}{c} e^{2i k_X y \sin \theta}\\
\bm E_1\cdot \bm B_2^* &=& \frac{E_X^2}{c} \cos^2\theta \ e^{2i k_X y \sin \theta}
\end{eqnarray}

The signal contributions from diagrams $D_1$ is thus:
\begin{eqnarray}
P_{D_1}(t, \tau) &=& P^{\mu\mu}_{D_1}(t, \tau) +P^{\mu m}_{D_1}(t, \tau) +P^{m\mu}_{D_1}(t, \tau)\nonumber\\
&=& 0 +\frac{1}{9c}\frac{i}{\hbar^3}
\langle \bm \mu_L\cdot\bm \mu_R\rangle
U(\tau)
\Big(\langle\bm \mu_R\cdot\bm m_L\rangle + \cos^2\theta \langle\bm m_R\cdot\bm \mu_L\rangle\Big)
 \bm E_3 E_X^2 \cos \theta\nonumber\\
 &=& 
\frac{1+\cos^2\theta}{9c}\frac{i}{\hbar^3} \langle \bm \mu_L\cdot\bm \mu_R\rangle
U(\tau)
\langle\bm \mu_R\cdot\bm m_L\rangle \bm E_3 I_X
\label{eq:pd1imp}
\end{eqnarray}
\noindent where $I_X = E_X^2$ is the intensity of the X-ray pulses. In Eq. \ref{eq:pd1imp}, we have used that $\langle\bm \mu_R\cdot\bm m_L\rangle =\langle\bm m_R\cdot\bm \mu_L\rangle$ for linear absorption terms.

Hence, the pseudo-scalar $\bm \mu\cdot \bm m$ appears in the leading term to the cross-polarized XTG signal.
It is closely related to the rotatory strength given by $R_{cg}=\text{Im}(\bm \mu_{gc}\cdot\bm m_{cg})$ where $g$ and $c$ are the electronic ground and core-excited states respectively.
The rotatory strength $R_{cg}$ is closely related to the differential absorption between left and right CPL measured in standard CD measurement\cite{warnke2012circular} by:
\begin{equation}
R_{cg} \approx \frac{3\ln 10}{16\pi^2cN_a}\int_{\omega_{cg}-\epsilon}^{\omega_{cg}+\epsilon} d\omega \frac{\epsilon_L(\omega) - \epsilon_R(\omega)}{\omega}
\end{equation}
\noindent where $\epsilon_L$ and $\epsilon_R$ are the molar extinction coefficients measured with left and right CPL respectively. The measured XCD signal is given by $S_\text{CD}(\omega) = A_L(\omega) - A_R(\omega)$ where $A(\omega) = \epsilon(\omega) cl$ is the absorbance, $c$ the sample concentration and $l$ the interaction length.

Diagrams $D_1$ to $D_4$ give the same contribution within the approximations discussed above, and the nonlinear polarization $P_\text{XTG}^{(3)}$, Eq. \ref{eq:sumPdi}, is 4 times the result given in Eq. \ref{eq:pd1imp}.

When the spectral bandwidth of the incoming pump is accounted for, see Appendix \ref{app:resonance}, the cross polarization XTG signal is given by:
\begin{eqnarray}
 S_\text{XTG}(\tau, \omega_p) &=& \Big|4 \frac{1+\cos^2\theta}{9\hbar^3 c} 
\langle \bm \mu_L \mathcal G(\omega_\text{pr})\bm \mu_R\rangle U(\tau)
\langle \bm \mu_R \mathcal G(\omega_p)\bm m_L\rangle 
\bm E_3 I_X \Big|^2
\label{eq:xcdxtgsimple}\\
 &\propto&  |S_\text{CD}(\omega_p)|^2   
\end{eqnarray}

\noindent where $\omega_p$ is the pump central frequencies and $\omega_\text{pr}$ is the probe central frequency.
The delay $\tau$ change the prefactor of the signal through the classical propagator $U(\tau)$ and is independent from the chiral observable.
By scanning $\omega_p$ at a fixed delay $\tau$, one can reconstruct the amplitude squared of the XCD signal. 
By changing $\omega_p$, the phase matching condition is modified and one must use a detector able to follow the corresponding angular change, use a sufficiently large array detector or account for the losses due to partial phase matching.
We stress the fact that while cross-polarization XTG is sensitive to chirality, i.e. it vanishes in achiral samples, it is unable to discriminate opposite enantiomers since the rotatory strength is squared.
Heterodyne-detected XTG could be the key to discriminate enantiomers by measuring the signal phase.

If the polarization of the two pumps is not perfectly crossed or if their intensity is not equal, achiral contributions would appear as potential spurious signal. For example, if we consider a more general elliptic polarization for $\bm E_1$:
\begin{equation}
\bm E_1 = E_{1x}\bm e_x + E_{1y}e^{i\varphi}\bm e_y 
= \frac{1}{\sqrt{2}}(E_{1x}-iE_{1y}e^{i\varphi})\bm e_L + \frac{1}{\sqrt{2}}(E_{1x}+iE_{1y}e^{i\varphi})\bm e_R 
\end{equation}
\noindent then the dot product for the achiral contribution would not vanish anymore:
\begin{eqnarray}
\bm E_1\cdot \bm E_2^* &=& \frac{E_2}{2}(-i)\cos\theta e^{2ik_x y \sin\theta}\Big(E_{1x}(\bm e_L+\bm e_R)\cdot(\bm e_L-\bm e_R)+E_{1y}e^{i\varphi}(\bm e_L-\bm e_R)\cdot(\bm e_L-\bm e_R)\Big) \nonumber\\
&=& E_2E_{1y}e^{i\varphi} \cos\theta e^{2ik_x y \sin\theta}
\end{eqnarray}

\section{Application to the chiral response of L-tryptophan}

\begin{figure}[h]
    \centering
    \includegraphics[width=0.5\textwidth]{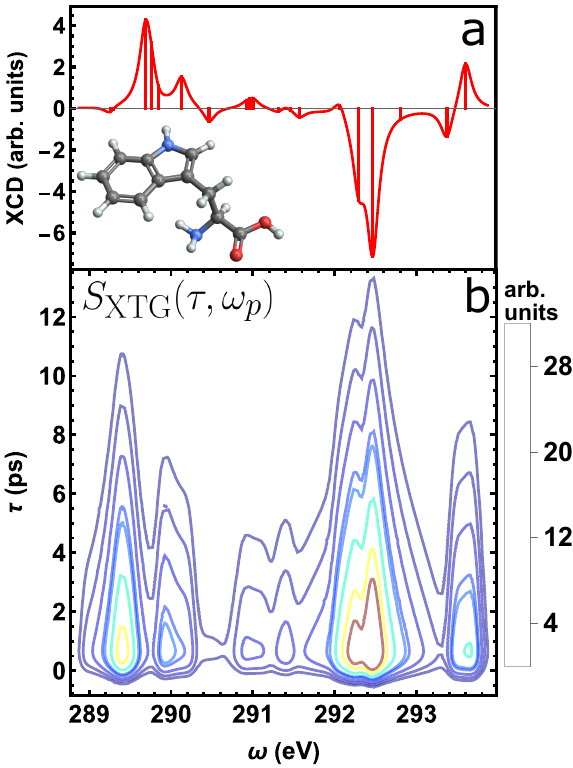}
    \caption{a) XCD signal of L-tryptophan (structure in insert) at the carbon K-edge. b) Cross-polarized XTG signal on the same molecule.}
    \label{fig:signal}
\end{figure}

We have calculated the cross-polarized XTG signal for a randomly oriented ensemble of L-tryptophan (Fig. \ref{fig:signal}a insert) with an X-ray incident pulse tuned at the C K-edge ($\omega_p$=288.5 - 294 eV).
The geometry was optimized at the HF/cc-pVDZ level of theory using the Molpro program package\cite{werner2020molpro}.

The core excited states are then calculated in separate RASSCF calculations, by freezing the optimization of the 1s core orbitals of each C atom, rotating them into the active space and restricting their occupation to a single electron. 
Computations were done at the RASSCF(9,7)/cc-pVDZ level of theory. The active space contained the singly occupied 1s orbital of the chosen C atom and valence orbitals ranging from HOMO-3 to LUMO+1.

In Fig. \ref{fig:signal}, the XCD and cross-polarized XTG signals are shown in panels a and b respectively. The XCD signal is computed using $S_\text{XCD}(\omega_p) = -\text{Im}\langle \bm \mu_R \mathcal G(\omega_1)\bm m_L\rangle$ where $\langle \bm \mu_R \mathcal G(\omega_1)\bm m_L\rangle$ is given in Eq.\ref{eq:xcdcorfunc}.
The stick spectrum of the transitions is shown in Fig. \ref{fig:signal}a and a phenomenological lifetime broadening for the C 1s excited of 0.1 eV was applied to display the spectrum\cite{nicolas2012lifetime}.
The cross-polarized XTG has been computed using Eq.\ref{eq:xcdxtgsimple}. A simple classical propagator of the form:
\begin{equation}
    U(\tau) = \frac{1}{2}\Big(1+\text{erf}(\frac{\tau}{t_r})\Big) e^{-t/t_d}
\end{equation}
\noindent where $t_r = 0.5$ ps and $t_d=5$ ps accounts for the instrument response function and for the classical relaxation time respectively. Typically, $t_d$ describes a thermal decay, by diffusion or vibrational relaxation.
As seen in Fig.\ref{fig:signal}b, horizontal slices of the cross-polarized XTG signal recover the amplitude square of the XCD signal.
Systems with short dephasing time and long thermal relaxation would allow for a larger time-window to detect XCD using the XTG signal.

\section{Conclusion}

In this study, we showcase that homodyne-detected cross-polarized X-ray Transient Grating is a remarkably sensitive technique for probing molecular chirality.
In the optical case, as discussed by Terazima\cite{terazima1995new}, the dephasing time can exhibit a relatively prolonged duration. 
This characteristic imposes limitations on the technique, particularly for longer delay times (hundreds of ps), where the transient grating could already show a significant decay.
In scenarios where core-holes are excited, the dephasing time is notably shorter than the core lifetime, typically falling within the range of approximately of few femtoseconds. This introduces the possibility of factorizing the X-ray Transient Grating (XTG) matter correlation function at very short times.
In the case of pulses lasting a few tens of femtoseconds, the phenomenon of decoherence can manifest within the pulse duration itself. This implies that factorization is consistently possible, and the signal effectively probes rotatory strengths even in scenarios where no delays are introduced between the pump and the probe pulses.
If this hypothesis proves to be accurate, it would open the door to probing chirality using self-diffraction. In the self-diffraction scenario, two pulses intersect on the sample, and one of the two pumps undergoes diffraction by the transient grating that is forming. This corresponds to a degenerate XTG at zero delay (the probe pulse is one of the pump pulses)\cite{morillo2023all}.

If the cross-polarization XTG experiment with attosecond pulses was pursued, it could be expected that the full four-point correlation function would have to be considered fully in that case.
Deviation from the XCD measurement could in turn be considered as a signature of the core-hole decoherence process.
Finally, if the polarization of the emitted electric field $\bm E_s$ were to be resolved, it would be possible to consider cross-polarization contributions from $\bm E_3$ and $\bm E_s$, allowing the detection of rotatory strength after the delay $\tau$.

\appendix
\section{Spectral measurement by frequency scanning}
\label{app:resonance}

In this appendix, we show how the spectral features of XCD are accounted for when the central frequency $\omega_p$ of the pump pulses is scanned.
Starting from Eq. \ref{eq:pd1mum} for $P_{D_1}^{\mu m}(t, \tau, \omega_p)$ and Fourier transforming the pulse $\bm E_1$, we get
\begin{multline}
P_{D_1}^{\mu m}(t, \tau,\omega_p) = \Big(-\frac{i}{\hbar}\Big)^3\int dt_3 dt_2dt_1 \frac{d\omega_1}{2\pi}
\langle \bm \mu_L \mathcal G(t_3)\bm \mu_R \mathcal G(t_2)\bm \mu_R \mathcal G(t_1)\bm m_L\rangle \\
\times
\bm E_3(t-t_3)\bm E_2^*(t-t_3-t_2+\tau)\bm B_1(\omega_1)e^{-i\omega_1(t-t_3-t_2-t_1+\tau)}
\label{eq:pd1app1}
\end{multline}
\noindent where $\bm B_1(\omega) = \int dt \bm B_1(t) e^{i\omega t}$. The Fourier transform over the propagator $\mathcal G(t_1)$ is given by:
\begin{equation}
\mathcal G(\omega_1) = \sum_{c} \frac{|cg\rangle\rangle\langle\langle cg|}{\omega - \omega_{cg}+i\Gamma_{cg}}
\end{equation}
\noindent where the Liouville space double bracket notation\cite{mukamel1995principles} has been used. $c$ and $g$ represent the core-excited states and the ground state respectively, and $\Gamma_{cg}$ are phenomenological broadening rates for core-excited states.

When the pump pulses spectral bandwidth is narrow compared to the core excited state lineshape, we can approximate the spectral envelope by a Dirac delta function centered at the pump frequencies $\omega_p$: $\bm B_1(\omega_1) = \bm B_1 \delta(\omega_1 - \omega_p)$. 
The temporal envelope of $\bm E_2$, which has the same spectral profile as $\bm E_1$, is short compared to the classical dynamics following the core hole dephasing time and can still be kept as impulsive in time domain.
Finally, the temporal envelope of the probe pulse is independent of that process and is taken as impulsive for simplicity.
Under these approximations, Eq. \ref{eq:pd1app1} becomes:
\begin{multline}
P_{D_1}^{\mu m}(t, \tau,\omega_p) = \Big(-\frac{i}{\hbar}\Big)^3\int dt_2 \frac{d\omega_1}{2\pi}
\langle \bm \mu_L \mathcal G(t)\bm \mu_R \mathcal G(t_2)\bm \mu_R \mathcal G(\omega_1)\bm m_L\rangle 
\bm E_3\bm E_2^*(-t_2+\tau)\bm B_1(\omega_1)e^{i\omega_1(t_2-\tau)}\\
=  \Big(-\frac{i}{\hbar}\Big)^3 \bm E_3\bm E_2^*\bm B_1 \langle \bm \mu_L \mathcal G(t)\bm \mu_R\rangle U(\tau) \langle \bm \mu_R \mathcal G(\omega_1)\bm m_L\rangle 
\label{eq:pd1app2}
\end{multline}
\noindent where 
\begin{equation}
\langle \bm \mu_R \mathcal G(\omega_p)\bm m_L\rangle
= \sum_c \frac{\bm \mu_{gc}\cdot\bm m_{cg}}{\omega_p - \omega_{cg}+i\Gamma_{cg}}
\label{eq:xcdcorfunc}
\end{equation}

\section*{Acknowledgements}
\addcontentsline{toc}{section}{Acknowledgements}

Work by JRR was supported by the U.S. Department of Energy  (DOE), Office of Science, Basic Energy Science  (BES), Chemical Sciences, Geosciences and Biosciences Division (CSGB) under Contract No. DE-AC02-06CH11357.

\bibliographystyle{unsrt}
\bibliography{biblio}

\end{document}